# Convex Combination of Overlap-Save Frequency-Domain Adaptive Filters

Sihai GUAN [1], Zhi LI [1]

**Abstract**: In order to decrease the steady-state error and reduce the computational complexity and increase the ability to identify a large unknown system, a convex combination of overlap-save frequency-domain adaptive filters (COSFDAF) algorithm is proposed. From the articles available, most papers discuss convex combinations of adaptive-filter algorithms focusing on the time domain. Those algorithms show better performances in convergence speed and steady-state error. The major defect of those algorithms, however, is the computational complexity. To deal with this problem and motivated by frequency-domain adaptive filters (FDAF) and convex optimization, this paper gives an adaptive filter algorithm, that consists of combining the two FDAFs using the convex combination principles and derives a formula to update the mixing parameter. The computational complexity of the COSFDAF is analyzed theoretically. The simulation results show that no matter what kinds of signal to be processed, whether correlated (i.e. colored noise) or uncorrelated (i.e. white noise), the proposed algorithm has better performance in identify the unknown coefficients when compared to a single overlap-save FDAF or the convex combination of two time-domain adaptive filters.

## I. INTRODUCTION

The least mean square (LMS) algorithm has probably become the most popular algorithm for its simple configuration, low computational complexity, effective tracking capability and easiness of implementation when used in a time-varying environment. For noise eliminate, interference suppression, or channel/plant identification, adaptation step serves to select the exact balance between convergence speed and low residual misadjustment. To resolve this contradiction, a combination of one fast and one slow LMS adaptive filter was proposed [1], are the output of this filter can produce an overall output of improved quality. Recently, convex combination of adaptive filters enjoys much popularity [2][3][4]. Its development is motivated by the idea of combining the performance of different adaptive filters to offer complementary capabilities. The mean-square performance of the convex combination of LMS has been analyzed in [5]. However, when the filter taps in large, the computational complexity has been plagued by the adaptive filter algorithms. Apart from convex combination, there is the affine combination method [6]. Because of potentially savings in the computational complexity and the DFT and the filter bank structures generate signals that are approximately orthogonal, the thought of FDAF algorithm was first proposed in [7]. Compared to the adaptive filter algorithm in the time domain, the FDAF algorithm shows better performance in computational complexity. Duel to use of the discrete fourier transform (DFT) and filter bank structures to generate signals that are approximately orthogonal [8]. The FDAF algorithm has been widely used [9][10][11]. In addition, subband adaptive filter, convex combination of subband adaptive filters is offered in [12][13]. Major drawback of the adaptive algorithm is how to decrease the steady-state error. After analysis of the above, in order to reduce the steady state error and decrease computational complexity, this paper propose a new adaptive filter algorithm that consists of two FDAF, use the convex combination principles and derives a formula to update the mixing parameter based on a gradient descent method. In addition, implementation of the algorithm is obtained based on overlap-save. We call our proposed algorithm the convex overlap-save FDAF (COSFDAF). It is noteworthy that large plant identification based on convex combinations of FDAFS there is rarely reported. In addition, a detailed realization of the COSFDAF algorithm, computational complexity, and numerical simulation is given in this paper. For simplicity, we assume that the input signal is real.

✉ Zhi Li
15107739422@163.com
Sihai Guan
gcihey@sina.cn

[1] School of Electro-Mechanical Engineering Xidian University, No. 2 South Taibai Road, Xi'an, Shaanxi, 710071, China

## II. Proposed COSFDAF Algorithm

Both circular convolution and overlap-save can achieve convolution using the FFT. Compared to circular convolution, overlap-save needs a gradient constraint, and uses $2M$-point DFTs. It performs a linear convolution between a finite-length sequence and an infinite-length sequence by appropriately partitioning the data. The overlap-save FDAF algorithm was proposed in [8]. In this section, we formulate the COSFDAF algorithm. The scheme of the COSFDAF algorithm in the $k$th block (with $k = 1, 2, \cdots, K$) is shown in Fig. 1, where some variables appear. Each variable is divided into $K$ groups, $\mathbf{x} = [\mathbf{x}_1, \mathbf{x}_2, ..., \mathbf{x}_k, ..., \mathbf{x}_K]$, wherein $\mathbf{x}_k = [x(M+k), ..., x(kM)]$, $\mathbf{X}_k = \text{FFT}[\mathbf{x}_k]$ is a $2M$-point Fourier transform of $\mathbf{x}_k$, $\mathbf{d} = [\mathbf{d}_1, \mathbf{d}_2, ..., \mathbf{d}_k, ..., \mathbf{d}_K]$, wherein $\mathbf{d}_k = [d(M+k), ..., d(kM)]$, $\mathbf{D}_k = \text{FFT}[\mathbf{d}_k]$ is a $2M$-point Fourier transform of $\mathbf{d}_k$, $\mathbf{y} = [\mathbf{y}_1, \mathbf{y}_2, ..., \mathbf{y}_k, ..., \mathbf{y}_K]$, wherein $\mathbf{y}_k = [y(M+k), ..., y(kM)]$, and $\mathbf{Y}_k = \text{FFT}[\mathbf{y}_k]$ is a $2M$-point Fourier transform of $\mathbf{y}_k$. Each component is adapted with its own rules and error, while the mixing parameter $\lambda(k)$ is chosen to minimize the quadratic of the frequency-domain error of the overall filter.

Based on Fig.1, the output of the COSFDAF is expressed as

$$\mathbf{y}_k = \lambda(k)\mathbf{y}_{1,k} + [1-\lambda(k)]\mathbf{y}_{2,k}, \quad 0 \leq \lambda(k) \leq 1 \tag{1}$$

where $\mathbf{y}_{1,k} = [y_{1,k}(1), y_{1,k}(2), ..., y_{1,k}(M)]$ and $\mathbf{y}_{2,k} = [y_{2,k}(1), y_{2,k}(2), ..., y_{2,k}(M)]$.

The relationship between the equivalent filter coefficient and the two sub-filters is shown as

$$\mathbf{W}_k = \lambda(k)\mathbf{W}_{1,k} + [1-\lambda(k)]\mathbf{W}_{2,k} \tag{2}$$

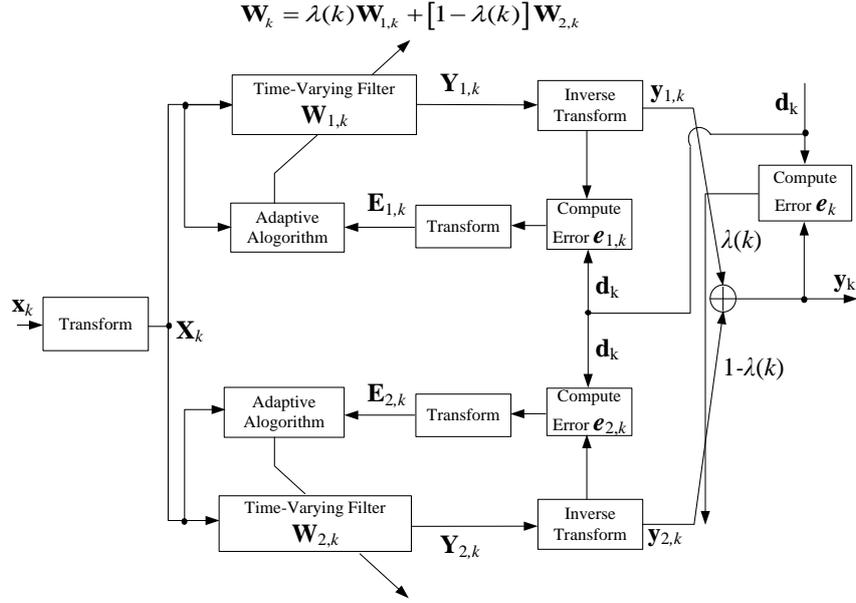

Fig.1 COSFDAF scheme in the $k$th block time.

For adaptation of the mixing parameter $\lambda(k)$, we will adapt a variable $a(k)$ using a sigmoidal function as $\lambda(k) = \mathrm{sgn}[a(k)]$.

The total error is described in the time domain.

$$\mathbf{e}_k = \mathbf{d}_k - \mathbf{y}_k \tag{3}$$

where $\mathbf{d}_k = [d_k(1), d_k(2), ...., d_k(M)]$ and $\mathbf{y}_k = [y_k(1), y_k(2), ...., y_k(M)]$.

The quadratic time-domain error can be calculated as

$$\mathbf{e}_k^2 = [\mathbf{d}_k - \mathbf{y}_k]^2 \tag{4}$$

Now we use the gradient descent method to minimize the quadratic error of the COSFDAF output in the $k$th block time, namely $\mathbf{e}_k^2$. Thus, the update equation for $a(k)$ is given as

$$\begin{aligned}
a(k+1) &= a(k) - \frac{\mu_a}{2} \frac{\partial \mathbf{e}_k^2}{\partial a(k)} \\
&= a(k) - \frac{\mu_a}{2} \frac{\partial \frac{1}{M} \sum_{m=1}^{M} e_k^2(m)}{\partial a(k)} \\
&= a(k) - \frac{\mu_a}{2M} \sum_{m=1}^{M} \frac{\partial e_k^2(m)}{\partial \lambda(k)} \frac{\partial \lambda(k)}{\partial a(k)} \\
&= a(k) - \frac{\mu_a}{M} \sum_{m=1}^{M} e_k(m) \left( y_{2k}(m) - y_{1k}(m) \right) \lambda(k)[1 - \lambda(k)]
\end{aligned} \quad (5)$$

For convenience, the complete COSFDAF algorithm is summarized in Table 1.

Table 1 The COSFDAF algorithm

Inputs

$\mu 1(0)$; $\mu 2(0)$; $\mu_a$; $\mu_{\max}$; $\beta$; $i=1,2$; $\alpha$; $\gamma_i$; $r$

Parameters Initialization

$a(0) = 0$; $\lambda(0) = 0$; $\mathbf{W}_i(0) = [0,0,...,0]^T_{1 \times M}$; $\mathbf{P}_i(0) = [1,1,...,1]_{M \times 1}$; $\mathbf{k} = [\mathbf{0}_M \ \mathbf{I}_M]$; $\mathbf{g} = \begin{bmatrix} \mathbf{I}_M & \mathbf{0}_M \\ \mathbf{0}_M & \mathbf{0}_M \end{bmatrix}$

Recursive Process

for $k = 1, 2, \cdots, K$

$\mathbf{d}_k = \left[ d(kM), ..., d(kM + M - 1) \right]^T$

$\mathbf{X}_k = \text{diag}\left\{ \text{FFT}\left[ x(kM - M), ..., x(kM + M - 1) \right]^T \right\}$

$\mathbf{Y}_{i,k} = \mathbf{X}_k \mathbf{W}_{i,k}$; $\mathbf{y}_{i,k} = \mathbf{k} \bullet \text{FFT}[\mathbf{Y}_{i,k}]$; $\mathbf{e}_{i,k} = \mathbf{d}_k - \mathbf{y}_{i,k}$; $\mathbf{E}_{i,k} = \text{FFT}\left[\mathbf{k}^T \mathbf{e}_{i,k}\right]$

$\mathbf{P}_{i,k} = \gamma_i \mathbf{P}_i(0) + (1 - \gamma_i) |\mathbf{X}_k|^2$

$\mathbf{W}_{i,k+1} = \mathbf{W}_{i,k} + 2 \text{FFT}\left\{ \mathbf{g} \bullet \text{IFFT}\left[ \mu_i(k) \mathbf{X}_k^H \mathbf{E}_{i,k} \frac{1}{\mathbf{P}_{i,k}} \right] \right\}$

$\lambda(k) = \text{sgn}[a(k)] = \frac{1}{1 + \exp[-a(k)]}$

$\mathbf{Y}_k = \lambda(k) \mathbf{Y}_{1,k} + [1 - \lambda(k)] \mathbf{Y}_{2,k}$

$\mathbf{W}_k = \lambda(k) \mathbf{W}_{1,k} + [1 - \lambda(k)] \mathbf{W}_{2,k}$

$\mathbf{y}_k = \mathbf{k} \bullet \text{FFT}[\mathbf{Y}_k]$; $\mathbf{e}_k = \mathbf{d}_k - \mathbf{y}_k$

$a(k+1) = a(k) - \frac{\mu_a}{M} \sum_{m=1}^{M} e_k(m) [y_{2k}(m) - y_{1k}(m)] \lambda(k)[1 - \lambda(k)]$

end

In Table 1, parameters appear that deserve to be explained, $\mu_i$, $i=1,2$ are used to denote the step sizes of the two LMS filters, $\beta$ is a threshold close to 1, $\alpha$ is the endpoint, $\gamma$ is a forgetting factor, $r$ is used to get the new value of the step sizes.

For $\lambda(k)$ in Table 1, some constraint conditions have to be added. Those constrains are summarized in Table 2.

Table 2 Constrains for $\lambda(k)$

if $\lambda(k+1) < 1 - \beta$

$$\mu_1(k+1) = \frac{\mu_1(k)}{r}; \quad \mu_2(k+1) = \frac{\mu_2(k)}{r}; \quad \mathbf{W}_{1,k+1} = \mathbf{W}_{2,k+1}; \quad a(k+1) = 0; \quad \lambda(k+1) = 0.5$$

end

if $\left[\lambda(k+1) > \beta \text{ and } r\mu_1(k) < \mu_{\max}\right]$

$$\mu_1(k+1) = r\mu_1(k); \quad \mu_2(k+1) = r\mu_2(k); \quad \mathbf{W}_{1,k+1} = \mathbf{W}_{2,k+1};$$
$$a(k+1) = 0; \quad \lambda(k+1) = 0.5$$

end

if $\left[\lambda(k+1) > \beta \text{ and } r\mu_1(k) > \mu_{\max}\right]$

$$a(k+1) = a^+; \quad \lambda(k+1) = \beta$$

end

## III. COMPUTATIONAL COMPLEXITY

The computational complexity of the COSFDAF algorithm is summarized for comparison with the convex variable step-size LMS CVSLMS [5] algorithm. The total number of multiplications for each implementation provided reasonably accurate comparative estimates of their overall complexity.

For the overlap-save FDAF, the total number of real multiplications is $10M\log_2(2M) + 16M$ [8]. As the COSFDAF algorithm combines two overlap-save FDAF filters, it needs $20M\log_2(2M) + 32M$ real multiplications for every $M$th output sample. Including the computational burden of computing its mixing parameter, the COSFDAF algorithm requires $3M+1$ real multiplications to compute the filter output and to update $a(k)$. Because the LMS algorithm with $M$ real weights requires $M$ multiplications to compute its output and another $M$ multiplication to update the weight vector, a total of $2M+3$ real multiplications to needed to produce each output sample [16]. So, $2M^2+3M$ real multiplications are required for every $M$th output sample. As the CVSLMS algorithm combines two LMS filters, it needs $4M^2+6M$ real multiplications for every $M$th output sample. Including the computational burden of computing its mixing parameter, the CVSLMS algorithm requires 4 real multiplications to compute the filter output and to update $a(k)$. For convenience, the computational complexity of the COSFDAF algorithm and the CVSLMS algorithms is summarized in Table 3.

Table 3 Computational complexity of COSFDAF and CVSLMS

| Algorithm | Multiplications |
|---|---|
| CVSLMS | $4M^2+6M+4$ |
| COSFDAF | $20M\log_2(2M)+35M+1$ |

From Table 3, some useful conclusions can be obtained. We see that for $M = 32$, the proposed algorithm and the CVSLMS algorithm already have the same computational complexity. To more filter size $M$ increase, more complex unknown system, the computational load ratio decease, so the COSFDAF algorithm has a considerable complexity advantage. The dramatic reduction in the number of multiplications is obtained because of the block updating and the DFT algorithm [8]. Thus, the COSFDAF algorithm is more efficient for satisfied with a large number of taps.

## IV. SIMULATION ANALYSIS

In order to further analyses the COSFDAF algorithm, two sets of experiments are designed with aims: Case 1 are shown that with an uncorrelated input signal of SNR = 20dB the performance of COSFDAF is better than a single overlap-save FDAF, and also better than CVSLMS [5] in identifying the unknown coefficients; Case 2 is similar to Case 1, but with a highly correlated input signal. For Case 1 and Case 2, curves of the excess mean-square error (EMSE) and the mean-square deviations (MSD) were exhibited.

Case 1: Uncorrelated input signal.

In the first part, we want to compare the performance of COSFDAF and a single overlap-save FDAF when the input signal is uncorrelated. 64000 iterations, 100 Monte Carlo, block size 64, forgetting factors $\gamma_1$ and $\gamma_2$ equal to 0.99, initial values $a(1)=0$, $\lambda(1)=0$, $\mu1 = 0.1$, $\mu2 = 0.008$, $\mu_{\max} = 4$, $\beta = 0.99$, $r = 4.5$, $a^+ = 4$, $\mu_a = 2000$, and $M$=64. Curves of EMSEs and MSDs are shown as Fig. 2.

Compared to the two single overlap-save FDAFs, the COSFDAF algorithm achieve comparable performance in terms of steady-state error and in identifying the unknown coefficients.

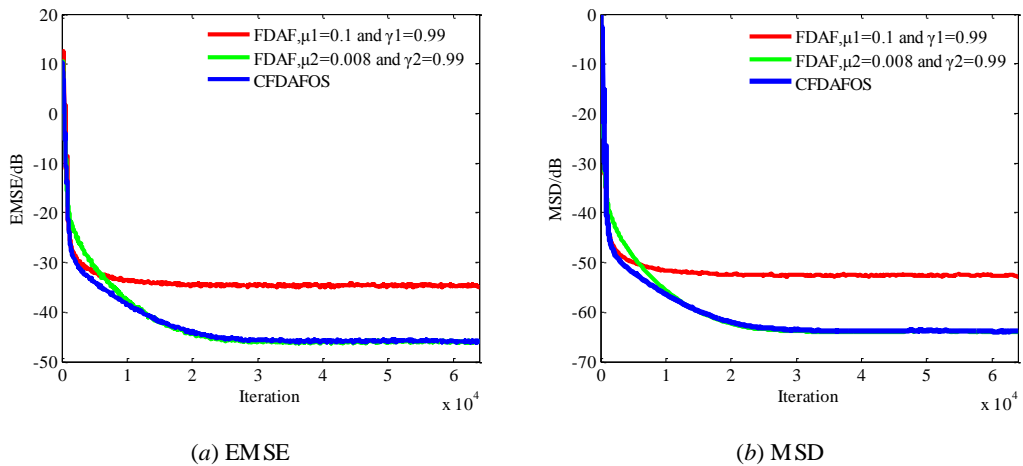

(a) EMSE   (b) MSD

Fig.2. Comparison of single overlap-save FDAFs and COSFDAF algorithms with uncorrelated input signal.

In the second part, we want to compare the performance of the COSFDAF and CVSLMS algorithms. The parameters set as: 128000 iterations. 50 Monte Carlo,

$M=32$, block size 32, forgetting factors $\gamma_1$ and $\gamma_2$ equal to 0.99, initial values $a(1)=0$ and $\lambda(1)=0.5$, $\mu1=0.1$, $\mu2=0.01$, $\mu_{max}=0.2$, $\beta=0.99$, $r=4.5$, $a^+=4$, and $\mu_a=100$.

From comparison of the final EMSEs in Fig. 3(*a*), the COSFDAF algorithm is lower than that of the CVSLMS algorithm by nearly 17 dB. Based on Fig. 3(*b*), the COSFDAF algorithm MSD is less than that of the CVSLMS algorithm by nearly 9 dB. Thus, compared to the CVSLMS algorithm, the COSFDAF algorithm achieves smaller steady-state error and better performance in identifying the unknown coefficients.

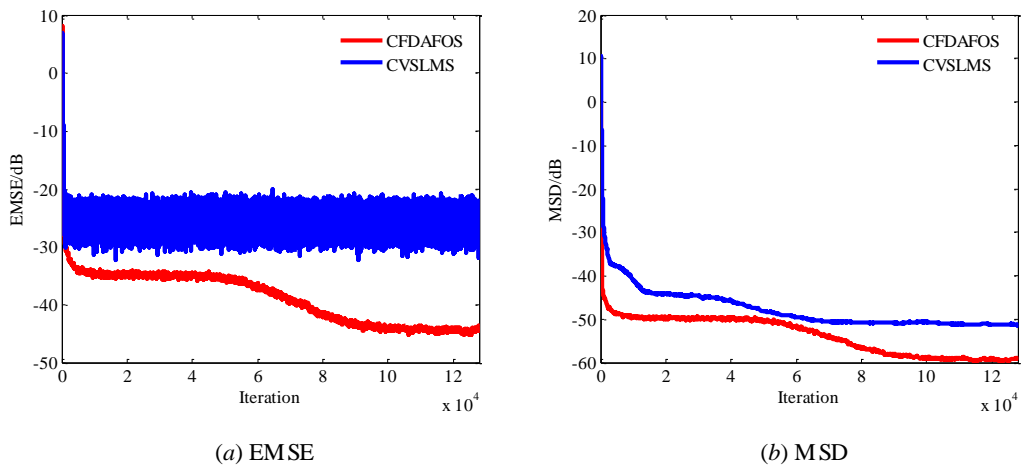

(*a*) EMSE  (*b*) MSD

Fig.3. Comparison of CVSLMS and COSFDAF algorithms with uncorrelated input signal.

Case 2: Highly correlated input signal.

The numerical simulation experiments are taken when the input signal correlation coefficient is 0.8. The parameter set as: 64000 iterations, 100 Monte Carlo, block size 64, forgetting factors $\gamma_1$ and $\gamma_2$ equal to 0.99, initial values $a(1)=0$, $\lambda(1)=0.5$, $\mu1=0.1$, $\mu2=0.008$, $\mu_{max}=4$, $\beta=0.99$, $r=4.5$, $a^+=4$, and $\mu_a=2000$, and $M=64$.

Compared to two single overlap-save FDAFs, the COSFDAF algorithm achieve comparable performance in terms of steady-state error and in identifying the unknown coefficients.

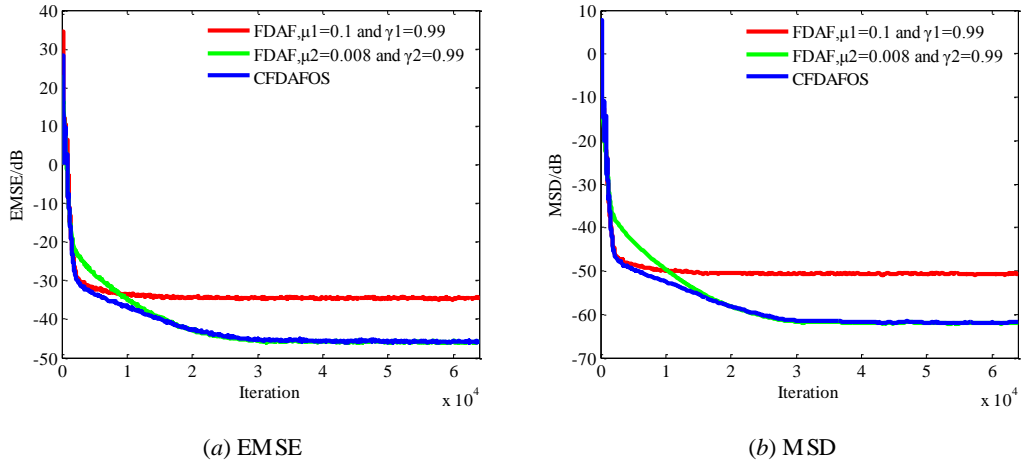

(a) EMSE  (b) MSD

Fig.4. Performance of single overlap-save FDAFs and COSFDAF algorithms with correlated input signal.

Finally, we compare the performance of the COSFDAF and CVSLMS algorithm whit input signal is highly correlated. The parameters set as: 128000 iterations, 50 Monte Carlo, and $M=32$, The block size 32, forgetting factors $\gamma_1$ and $\gamma_2$ equal to 0.99, initial values $a(1)=0$ and $\lambda(1)=0.5$, $\mu_1 = 0.1$, $\mu_2 = 0.01$, $\mu_{max} = 0.2$, $\beta = 0.99$, $r = 4.5$, $a^+ = 4$, and $\mu_a = 100$.

The EMSEs and MSDs are shown in Fig. 5. Based on Fig. 5(a), the COSFDAF algorithm EMSE is less than that of the CVSLMS algorithm by nearly 19 dB, and the COSFDAF algorithm final MSD is less than that of the CVSLMS algorithm by nearly 10 dB. Thus, compared to the CVSLMS algorithm, the COSFDAF algorithm achieves better performance in identifying the unknown coefficients of the large adaptive filter system.

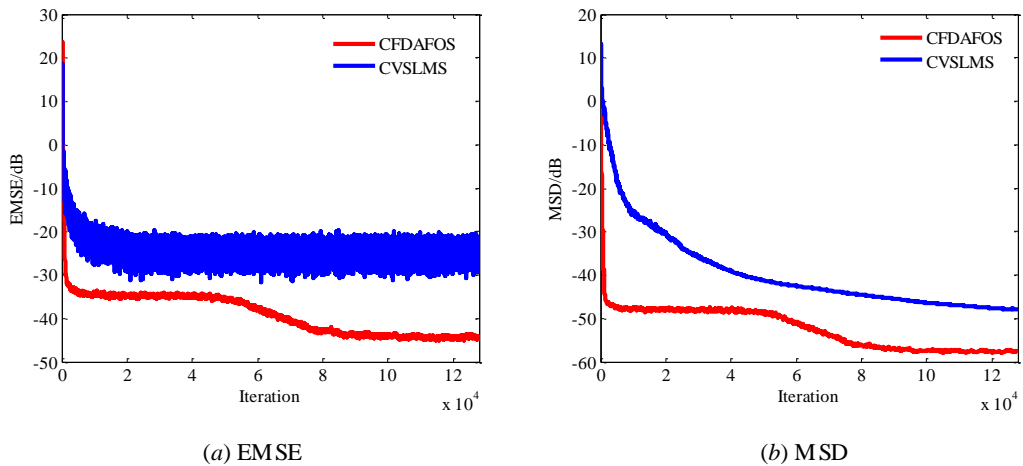

(a) EMSE  (b) MSD

Fig.5. Performance of CVSLMS and COSFDAF algorithms with correlated input signal.

The above analysis and comparison of available simulations can be shown as follows. Whether the input signal is uncorrelated or highly correlated, the COSFDAF algorithm obtain smaller steady-state error and have better performance in identifying the unknown system when compared to single overlap-save FDAF. Moreover, regardless of input signal correlation, COSFDAF have better performance in identifying the unknown coefficients of a large adaptive filter system when compared to the CVSLMS algorithm. The proposed algorithm and the CVSLMS algorithm already have the same computational complexity, and when the filter size *M* to larger, the COSFDAF algorithm has a considerable complexity advantage over the CVSLMS algorithm.

## V. CONCLUSION

In this paper, a novel convex combination of overlap-save frequency-domain adaptive filters is proposed. The COSFDAF algorithm operates in the frequency-domain, and consists of two FDAFs with convex combination, and the mixing factor $\lambda(k)$ of the recursive formula is derived. The computational complexity ratio between the COSFDAF algorithm and the CVSLMS algorithm is analyzed, theoretically, and significantly reduced for large filter size. The analysis and simulation results show that the new algorithm has better performance in term of computational complexity and steady-state error when compared to the single overlap-save FDAF. Moreover, whether the input signal is highly correlated or not, the algorithm has better performance than the CVSLMS algorithm in identifying the unknown coefficients of a large adaptive filter system. So the COSFDAF algorithm provides a new method to solve the trade-off between steady-state error and computational complexity. Therefore, the proposed algorithm has great potential value for plant identification with a large number of taps.

**Acknowledgments** This work was supported by National Natural Science Foundation of China (61074120, 61673310) and the State Key Laboratory of Intelligent Control and Decision of Complex Systems of Beijing Institute of Technology.## REFERENCES

[1]. Martinez-Ramon M, Arenas-Garcia J, Navia-Vázquez A, "An adaptive combination of adaptive filters for plant identification," *14th Int. Conf. Digital Signal Process.. DSP 2002. IEEE*, pp. 1195-1198, 2002.
[2]. Shi L, Lin Y, "Convex combination of adaptive filters under the maximum correntropy criterion in impulsive interference," *IEEE Signal Process. Letters*, vol. 21, pp. 1385-1388, Nov. 2014.
[3]. Sayed A H, "Adaptive networks," *Proceedings of the IEEE*, vol. 102, pp. 460-497, April 2014.
[4]. Ferrer M, Gonzalez A, de Diego M, "Convex combination filtered-x algorithms for active noise control systems" *IEEE Trans. Audio, Speech, and Lang. Process.*, ,vol. 21, pp. 156-167, June 2013.
[5]. Arenas-García J, Figueiras-Vidal A R, Sayed A H, "Mean-square performance of a convex combination of two adaptive filters," *IEEE Trans. Signal Process.*, vol.

54, pp:1078-1090, March 2006.
[6]. Bershad N J, Bermudez J C M, Tourneret J Y, "An affine combination of two LMS adaptive filters—transient mean-square analysis," *IEEE Trans. Signal Process.*, vol.56, pp.1853-1864, May 2008.
[7]. Dentino, M., McCool, J., Widrow, B., "Adaptive filtering in the Frequency-Domain," *Proceedings of the IEEE*, vol. 66, pp. 1658-1659, Dec. 1978.
[8]. Shynk J J, "Frequency-domain and multirate adaptive filtering," *IEEE Signal Process. Magazine*, vol. 9, pp.14-37, Jan. 1992.
[9]. Lee J, Huang H C, "On the step-size bounds of frequency-domain block LMS adaptive filters," *IEEE Signal Process. Letters*, vol. 20, pp.23-26, Jan. 2013.
[10]. Souden M, Benesty J, Affes S, "On optimal frequency-domain multichannel linear filtering for noise reduction," *IEEE Trans. Audio, Speech, and Lang. Process.*, vol. 18, pp. 260-276, Feb. 2010.
[11]. Guldenschuh M, De Callafon R, "Detection of secondary-path irregularities in active noise control headphones," *IEEE/ACM Trans. Audio, Speech and Lang. Process.*, vol. 22, pp.1148-1157, July 2014.
[12]. Sohn Sang-Wook, Yun J J, Lee J, "Convex combination of subband adaptive filters for sparse impulse response systems," *Midwest Symposium on Circuits and Systems,* pp.1-4, 2011.
[13]. Sohn S W, Lee J, Lee K P, "Subband adaptive convex combination of two NLMS based filters for sparse impulse response systems," *IEEE Statistical Signal Process. Workshop (SSP)*, 2012:201-204.
[14]. Arenas-Garcia J, "Mean-square performance of a convex combination of two adaptive filters," *IEEE Trans. Signal Process.*, vol. 54, pp.1078-1090, March 2006.
[15]. P. S. R. Diniz, Adaptive filtering, Fourth Edi. Boston, MA, USA: Springer US, 2013.
[16]. Mansour D, Gray A," Unconstrained frequency-domain adaptive filters," *IEEE Trans. Acoust. Speech & Signal Process.*, vol. 30,pp. 726-734, Oct. 1982.